\documentclass[preprint]{aastex}
\usepackage{array}
\usepackage{multicol}
\usepackage{amssymb}
\usepackage{epsfig}
\usepackage{ifthen}
\usepackage{rotating}

\newboolean{ispreprint}
\setboolean{ispreprint}{true}

\bibpunct[; ]{(}{)}{;}{a}{}{,}

\newsavebox{\thisbox}
\newlength{\thiswid}

\newcommand{\etal}{\textit{et al.}}

\shorttitle{ COSMOGRAPHY  AND COSMOLOGY: A BAYESIAN ANALYSIS}
\shortauthors{M.V. JOHN}

\begin{document}

\title{ COSMOGRAPHY, DECELERATING PAST,  AND COSMOLOGICAL MODELS: LEARNING THE BAYESIAN WAY}

\author{MONCY V. JOHN}

\affil{Department of Physics, St. Thomas College, Kozhencherri, Kerala
689641,
India; }
\email{moncy@iucaa.ernet.in}

\begin{abstract}

In this paper, using a significantly improved version of  the  model-independent, cosmographic approach to cosmology   (John, M. V. 2004, \apj, 614, 1), we  address an important question: Was there a decelerating past for the universe? To answer this,   the Bayes's probability  theory is employed, which is the most appropriate tool for quantifying our knowledge when it changes through the acquisition of new data. The cosmographic approach helps  to sort out   the models in which the universe was always accelerating  from those in which it decelerated for at least some time in the period of interest.  Bayesian model comparison technique is used to discriminate these rival hypotheses with the aid of recent releases of supernova data. We also attempt to provide and improve another example of Bayesian model comparison, performed between some Friedmann models, using the same data. Our conclusion, which is consistent with other approaches, is that the apparent magnitude-redshift data alone cannot discriminate these competing hypotheses. We also argue that the lessons learnt using Bayesian theory are extremely valuable to avoid frequent U-turns in cosmology.

\end{abstract}


\section{INTRODUCTION}

It is usual in cosmology that  when new data come in, we need to readjust  our parameter values or shall even attempt to modify our theoretical model itself. But each time while doing so, we are guided by our knowledge gathered so far about the universe. This kind of gradual learning is characteristic of sciences like cosmology, where one cannot conduct laboratory experiments.  A very interesting turn of events which unfolded in cosmology in the past ten years exemplifies this learning process. Till the early 1990s, a Friedmann cosmological model with either radiation or dust as energy component was conceived to be the standard model in cosmology and  it was generally believed that this standard model described the observed universe (at least from the time of nucleosynthesis) to a very good accuracy. Most of the widely discussed cosmological problems, which ought to have surfaced  in the classical epoch, were known to disappear when we incorporate into the standard model the theory of inflation, which envisages a flat  universe. The value of the Hubble parameter $H_0 \equiv 100 h$ km s$^{-1}$ Mpc$^{-1}$ quoted during that period was $0.4<h<1$  \citep{kolb}. There was no serious age problem and no real need for a cosmological constant $\Lambda$ in the late universe. The  beginning of the present phase of $\Lambda$-term or dark energy \citep{pad,ratra} in cosmology is the measurement of the Hubble parameter  reported in \cite{freedpi,pierce} and  summarized in \cite{jaffe}, that gave a more  specific  and high range $h>0.7$. This, along with flatness imposed by the theory of inflation created a short spell of  age problem,  which required a $\Lambda$-term for its solution. But a nonzero $\Lambda$ was always behind the curtains and hence its appearance was not conceived to be a major deviation in the standard model scenario. But then came the Type Ia supernovae (SNe Ia) data \citep{per99,rie98}, which required the presence of a $\Lambda$-term large enough to cause an accelerated expansion. The most recent release of SNe data \citep{riess1,perl1,riess04} prompts many cosmologists even to speculate that  some extremely unphysical energy densities, such as those with $w<-1$ in the equation of state $p=w\rho$, are required to explain the data. Though this change was gradual, what happened here is a U-turn from the standpoint of the early nineties, since the dynamics, energy content, equation of state, etc. of the universe have now become totally speculative. But this is typical of every learning process and cannot be termed unscientific. 

It has long been recognized that the application of Bayes's theorem in physical problems represents learning. [For a recent review, see \cite{dagostini}.] This theorem tells us how to adjust our plausibility assumptions regarding a hypothesis when our state of knowledge changes through the acquisition of new data.  Classical mathematicians such as Bernoulli, Bayes, Laplace and Gauss have found Bayes theorem useful in problems such as those in astronomy, thanks to its ability to learn. Later, because of difficulties with assigning prior probabilities (which were mistakenly considered to be purely subjective expressions of a person's opinions about hypotheses), Bayes's probability theory has gone out of favor in physical sciences  and was replaced by the more apparently objective `frequentist' approach. But the realization that the frequentist definition of probability is as subjective  as the Bayesian has called forth a re-examination of this controversy. If some simple desideratum such as `equivalent states of knowledge should be represented by equivalent probability assignments' (which is termed as Jaynes's consistency) is followed, the Bayesian approach will help to quantify the collective wisdom of scientists and hence can be made less subjective. When compared to the improper application of frequentist probability theory, the Bayesian approach is most powerful in   problems such as those in cosmology, where the process of learning and also the quantification and readjustment of plausibility assessments by the scientific community are very important.

The apparent magnitude-redshift ($m-z$) data of SNe Ia    are the only qualitative signature of an accelerated expansion and hence it is very important in understanding the dynamics of the universe. This accelerated expansion is  confirmed \citep{mvj}, provided there are no evolutionary effects for SNe \citep{drell}. For analyzing the data, we need an expression for the scale factor $a(t)$. If the attempt is  to obtain $a(t)$ as the solution of Einstein equation, one must know the energy densities present in the universe and also their equations of state. But here the data indicate that known forms of energies are unable to account for it. Under such circumstances,  the use of the traditional Friedmann solutions of the scale factor $a(t)$ obtained by assuming the presence of known energies in cosmological problems, and  particularly in the analysis of SNe data, can be very misleading.   Hence these SNe data are to be analyzed cosmographically, without making any specific assumptions on the energy densities in the universe. Such an analysis of SNe data in \cite{mvj} assumes only homogeneity and isotropy for the universe; i.e., the universe is assumed to have a Robertson-Walker (RW) metric. The  scale factor of the universe can most naturally be expanded into a  Taylor series in $t$ about the present epoch and we attempt to find its coefficients from observation, as it was done to evaluate the Hubble parameter and deceleration parameter in the original cosmographic approach \citep{wein}. Terms up to fifth-order were kept in the above series and it was assumed that they make a good approximation. The results of this and other  analyses of SNe data do not give  reason to believe that there were only standard model energy densities or even any other energies with equation of state of the form $p=w\rho$ in the present universe.  Instead, the likelihoods for the various expansion rates obtained in our calculation are very broadly peaked  and these indicate that there are a variety of choices for the energy densities in the present  epoch. The Taylor series expansion approach is extended to third-order and fourth-order in \cite{sahni} and \cite{visser}, respectively. Also \cite{q2,q3,q4}  have attempted similar model-independent analyses of SNe data.

When ensembles and repeated experiments are not possible, a natural and useful procedure in cosmology is to compare how best different models can account for the data,  using the Bayesian method. In  Bayesian model comparison, one computes odds ratios between different models. \cite{jaffe} and \cite{hobson} have  used Bayesian  theory to test the relative merits of different cosmological models. In one  such application of this technique, \cite{mvjvn} have compared the standard and inflationary models having nonzero $\Lambda$ with a new simple model having the scale factor $a(t) \propto t$.  This comparison  was made using the SNe data set in \cite{per99} and by assuming flat priors. Flat priors indicate that we are having no prior information regarding parameters of the models except that they lie in some fiducial range. In the present case, this is equivalent to stating that one depends only on the present SNe data for making the comparison. In \cite{mvjvn}, it was  found that the then available apparent magnitude-redshift data alone were not very much discriminatory between these different models.  However, to be true to the spirit of Bayesian theory,  our plausibility assignments should be updated  with the acquisition of new data. In this paper, we attempt to do this, by using the recent release of SNe data.

Another important question we try to answer in this paper is whether the data really endorse that   the universe was decelerating  in the past. For this,  a model with some decelerating phase in the past is compared with another model having no such phase, by using the Bayesian approach. In both cases, we assume the present universe to be accelerating.  It is claimed that this method of comparison is more robust than other investigations seeking evidence for a decelerating past since  both the Bayesian and cosmographic approaches mentioned above are used here. As mentioned earlier,   the scale factor is expanded into a fifth-order polynomial  in time (fifth-order is required for sufficient accuracy) and then the  combinations of various coefficients in this expansion  were separated into those which correspond only to acceleration during the entire period and those which had at least some decelerating phase in this period. Considering these as rival models, the Bayes factor is calculated. As an example of the Bayesian model comparison technique, we compare the    general relativistic and Newtonian explanations of deflection of light by sun, to show how powerful is the available observed data on deflection of light  in discriminating these explanations. We argue that, in contrast to this case,  the SNe data are not  capable of discriminating   the ``always accelerating" and ``decelerating in the past" cosmological models. The results have serious implications for modern theoretical cosmology for it is almost entirely built on the firm belief that the universe was decelerating in the past.

The paper is organized as follows. In section 2,  the general formalism of Bayesian model comparison employed here is given. The question of whether there was a decelerating past for the universe is discussed in section 3. Comparison of some Friedmann  cosmological models is discussed in section 4 and section 5 summarizes our conclusions.

\section{BAYESIAN MODEL COMPARISON}

The Bayes's theorem helps to evaluate the posterior (i.e., after analyzing the data) probability $p(H_i|D,I)$ for a hypothesis $H_i$ given the data $D$ and the truth of some background information $I$, as

\begin{equation} \label{eq:prob}
p(H_i|D,I)= \frac {p(H_i|I)p(D|H_i,I)}{p(D|I)}.
\end{equation}
$ p(H_i|I) $ is called the prior (i.e., before analyzing the data) probability and is the probability for $H_i$, given the truth of $I$ alone. $p(D|H_i,I)$ is the probability for obtaining the data $D$ if the hypothesis $H_i$ and $I$ were true and is called the likelihood for the hypothesis. The factor in the denominator serves the purpose of normalization.

In Bayesian model comparison, one finds the odds ratios; i.e., the ratios between the posterior probabilities for different models. If we have to compare rival models $M_i$ and $M_j$, take the truths of these as the hypotheses $H_i$ and $H_j$, respectively and write, using Bayes's theorem,

\begin{equation}\label{eq:odds}
\frac{p(M_i|D,I) }{p(M_j|D,I)}= \frac {p(M_i|I)p(D|M_i,I)} {p(M_j|I)p(D|M_j,I)} \equiv O_{ij}.
\end{equation}

To some extent, evaluating the prior probability is subjective, since it depends only on the prior information $I$ and this may vary from person to person. But Bayesians view this theory as an attempt to quantify the collective wisdom of researchers working in the field and hence finding fiducial prior probabilities would be highly rewarding. However, if the information $I$ does not prefer one model over the other, the prior probabilities get cancel out and the odds ratio is simply
 
\begin{equation}\label{eq:bayes}
 O_{ij}= \frac {p(D|M_i,I)} {p(D|M_j,I)} \equiv B_{ij}.
\end{equation}

As mentioned above, the probability $p(D|M_i,I)$ for the data $D$, given that the model $M_i$ and $I$ are true,  is called the likelihood for the model $M_i$ and is denoted as ${\cal L}(M_i)$. For parameterized models, with parameters $\alpha , \beta , ..$,  this quantity can be evaluated as

\begin{equation}\label{eq:likelimod}
p(D|M_i,I)\equiv {\cal L}(M_i)=\int d\alpha \int d\beta ... p(\alpha , \beta , ...|M_i) {\cal L}_i (\alpha , \beta , ...),
\end{equation}
where $p(\alpha , \beta , ...|M_i)$ is the prior probability for the set of parameter values $\alpha , \beta , ..$ and ${\cal L}_i (\alpha , \beta , ...)$ is the likelihood for the combination. The latter quantity is often taken  to be 

\begin{equation}\label{eq:likelipar}
{\cal L}_i (\alpha , \beta , ...)=\exp \left[-\chi _i^2(\alpha , \beta , ..)/2 \right],
\end{equation}
where

\begin{equation}\label{eq:chi2}
\chi^2 = \Sigma _k \left( \frac{\hat{A}_k -A_k(\alpha , \beta , ..)}{\sigma_k}\right)^2
 \end{equation}
is the $\chi^2$-statistic. Here $\hat{A}_k$ are the measured values of the observable $A$, $A_k(\alpha , \beta , ..)$ are its expected values (from theory) and $\sigma _k$ are the uncertainties in the measurement of the observable.

In a certain volume $V$ of the parameter space, one can assign flat prior probability for all the parameters by taking $p(\alpha , \beta , ...|M_i) = $ constant throughout this volume. When normalized, this prior is simply $1/V$. In those special cases where there are no adjustable parameters in the model,  we have  $\delta$-function prior and from equations (\ref{eq:likelimod}) and (\ref{eq:likelipar}),

\begin{equation}\label{eq:likmod}
{\cal L}(M_i)=\exp (-\chi_i^2/2).
\end{equation}

$B_{ij}$ in equation (\ref{eq:bayes}) is referred to as the Bayes factor. The interpretation of this quantity is as follows \citep{drell}: If $1<B_{ij}<3$, there is evidence against model $M_j$, but it is not worth more than a bare mention. If $3<B_{ij}<20$, this evidence is positive.  If $20<B_{ij}<150$, it is strong and if $B_{ij} >150$, the evidence is very strong.

To appreciate the significance of this interpretation of the Bayes factor, let us compare the general relativistic and Newtonian explanations of the deflection of a light ray that just grazes the sun's surface. The theoretical prediction made by general relativity ($M_1$) in this case is $\theta _0 =1.75^{\prime \prime}$ whereas in the purely Newtonian case ($M_2$), it is  $\theta _0 =0.875^{\prime \prime}$. If we use the data $\hat {\theta}_0 =1.98\pm 0.16^{\prime \prime}$ obtained from the classic 1919 eclipse expedition to the island of Sobral \citep{dyson1,dyson2}, the Bayes factor calculated using equations (\ref{eq:bayes})-(\ref{eq:likmod}) above is $B_{12} \sim 10^{10}$. It may be noted that the $\chi^2$-values and hence the Bayes factor crucially depend on the error bars. Though the  error bars in the above case are now felt to be grossly underestimated, this published data took the world by storm in favor of GR in 1919 and the reason can be understood from the huge value of  $B_{12}$ we obtained above. If  all the 9 observational data points provided with error bars in Table 8.1 of \cite{wein} were used, one gets $B_{12} \sim 10^{82}$, a spectacularly large value for the Bayes factor to settle the issue in favor of general relativity. These kinds of results are common in tests of quantum theory too. These examples show that the above requirement $B_{ij} >150$ for an evidence to be very strong is really modest.

\section{ ACCELERATING VS. DECELERATING PAST}

We compare  models of an always accelerating universe  from time $t_0+T_{p}$  to the present ($t_0$ is the present time and $T_p$ is negative) with those which were decelerating for at least some time in this period. Computations were performed for various values of $T_{p}$ ranging from $-1\times 10^{17}$ s to $-5\times 10^{17}$ s, the absolute value of the latter  ($\approx 15$ Gyr) being usually quoted as the upper limit for the age of the universe. 

In \cite{mvj}, it was assumed that the scale factor of the universe can be approximated by a fifth-order polynomial  in time. With  the present value of the scale factor as $a_0$, the present deceleration parameter as $q_0$, and the present values of other parameters related to higher order derivatives as $r_0$, $s_0$ and $u_0$, this expression for $a(t)$ is  

\begin{eqnarray}
a(t_0+T)&=&a_0\left[
1+H_0T-\frac{q_0H_0^{2}}{2!}T^2+
\frac{r_0H_0^3}{3!}T^3-  \frac{s_0H_0^4}{4!}T^4 +\frac{u_0H_0^5}{5!}T^5
\right]  \nonumber \\
&\equiv& a_0\left[1+a_{(1)}T+a_{(2)}T^2+a_{(3)}T^3+a_{(4)}T^4+a_{(5)}T^5\right] 
\label{eq:a1}.
\end{eqnarray}

In addition to the above parameters, we have $k=0,\pm 1$, which is the curvature constant appearing in the RW metric and $M$, the absolute  luminosity of SNe.  The parameters $a_0$ and $M$ have reasonable flat priors for $ a_0>3000 $ Mpc and $-19.6<M<-19.1$ magnitudes, respectively. The upper bound for $a_0$ is chosen as 8000 Mpc,  large enough to incorporate spatially flat models. Consequently, $k=0$ need not be included in the calculations. In \cite{mvj}, marginal likelihoods were evaluated  for other parameters and found that the contributing ranges are $0.6<h<0.8$, $-2<q_0<1$, $-15<r_0<15$, $-65<s_0<65$, and $-150<u_0<150$. However, some combination of these parameter values were found to give no solution to the equation $1+z=a(t_0)/a(t_0+T)$ even for a time as past as $T=-10\times 10^{17}$ and  those values were  excluded. 

In the  Bayesian model comparison to find evidence for a decelerating past for the universe using new data, one should accept the above ranges as the prior information  obtained from the previous analysis. However,  a  modification is suggested to the effect that we shall begin by accepting  the present universe to be accelerating; i.e., $q_0 <0$. Considering the various other cosmological observations and the general perception among cosmologists, it is only reasonable to include this into the information $I$ we have, before analyzing the present data. The two hypotheses we want to compare may now be explicitly stated: (1) The universe is always accelerating from time $t_0+T_{p}$  to the present epoch (model $M_1$) and (2) There is at least one decelerating epoch for the universe during this period (model $M_2$). The factors $p(\alpha ,\beta , ..|M_i)$ in equation (\ref{eq:likelimod}), which are the prior probabilities for the parameters (given the truth of the respective models),  are taken in this case to be the flat probabilities $1/V_1$ and $1/V_2$ for models $M_1$ and $M_2$, respectively, where $V_1$ and $V_2$ are the volumes in the parameter space corresponding to each of them. For any particular combination of parameter values,  a sure test for the occurrence of deceleration  during $T_{p}<T<0$ is to plot 

\begin{equation}\label{eq:addot}
\ddot{a}(t_0 +T)=a_0[2a_{(2)}+6a_{(3)}T+12a_{(4)}T^2+20a_{(5)}T^3]
\end{equation}
for this interval and to see whether it becomes negative at any time during the period. The Bayes factor is then

\begin{equation}\label{eq:bayes1}
B_{12}=\frac{(1/V_1)\Sigma_k \int_{V_{1}}\exp(-\chi_1^2 /2) da_0 \;dh \; dq_0 \; dr_0 \; ds_0 \; du_0 \; dM}
{(1/V_2)\Sigma_k \int_{V_{2}}\exp(-\chi_2^2 /2) da_0 \;dh \; dq_0 \; dr_0 \; ds_0 \; du_0 \; dM}.
\end{equation}
Here, $\chi^2$ for a model is calculated using equation (\ref{eq:chi2}), with the replacement of $A$ by $m$, the apparent magnitude of the SN and the parameters $\alpha, \beta, ..$ are $a_0$, $h$, $q_0$,  $r_0$, $s_0$, $u_0$, $M$, and $k$. The  expression for $m$ to be used is 

\begin{equation}
m=5\log \frac{D}{1 \hbox {Mpc}}+25+M. \label{eq:m}
\end{equation}
$D/1$Mpc refers to the luminosity distance $D=r_1a_0(1+z)$, expressed in megaparsecs. The comoving coordinate $r_1$ can be found from 

\begin{equation}
\label{eq:r1}
\int_{T_{1}}^0 \frac{c\; dT}{a(t_0+T)}=\int_0^{r_1} \frac{dr}{1-kr^2}\equiv S_k^{-1}(r_1),
\end{equation}
where $t_1=t_0+T_1$ is the time at which an SN at $r_1$ emits the light and $S_k^{-1}(r_1)$ is equal to  $\sin^{-1} (r_1)$ for $k=+1$, and $\sinh^{-1} (r_1)$ for $k=-1$.

An important part of the calculation is the solution of the following equation, used to find $T_1$ in terms of $z$, for each combination of parameter values. This is done in a direct and purely numerical way [and differently from the way it was done in \cite{mvj}]:

\begin{equation}
 1+z=\frac{a(t_0)}{a(t_0+T_1)}= \frac{1}{1+a_{(1)}T_1+a_{(2)}T_1^2+a_{(3)}T_1^3+a_{(4)}T_1^4+a_{(5)}T_1^5}. \label{eq:numsolzT}
\end{equation} 
Another improvement is that this numerical solution for $T_1$ is found with the more reliable regula falsi method, rather than the Newton-Raphson method employed in the above paper. As a  consequence of these modifications,  $r_1$  has to be obtained by numerical integration in equation (\ref{eq:r1}).   Though it now requires more computation time, these changes have made the analysis free of the truncations that weakened the approximations
for certain parameters in the previous case. No point in the parameter space is now left out for the reason of breaking down of the approximations. As mentioned earlier, the only points left out are  those which do not have a solution for  equation (\ref{eq:numsolzT}) for all $z$ in the data set, even for the past $10^{18}$s. They are  not included in the volumes $V_1$ and $V_2$ either. The values of the Bayes factor obtained for various values of $T_{p}$ are tabulated in Table 1. The values obtained while using the 54 ``All SCP" SNe (data $D_1$) in \cite{perl1} (as reproduced in \cite{mvj}) and the 157 ``gold" data points (data $D_2$) in \cite{riess04}, respectively are given in two separate columns. It may also be noted that while using $D_2$, one does not have to include the parameter $M$ in the calculations since the data give $m-M$, the distance modulus in place of $m$. 

 The envelopes of the marginal likelihoods for parameters obtained in \cite{mvj} using $D_1$  were  found to be mostly unchanged. However, it can be seen that the afore-said modifications have made  significant improvement in the quality of the curves and for the purpose of verification, the new likelihoods for $h$, $q_0$, $r_0$, $s_0$, and $u_0$ are given here in Figures 1-5. The fluctuations have now disappeared and we have very smooth curves. These likelihoods were computed for  ranges wider than the ones in the previous case, but  the prior ranges for these parameters remained the same  (information $I$) in the computations of Bayes factors and other marginal likelihoods.

The results show that except in the case of using data $D_2$ for $T_p = -2\times 10^{17}$s, there is hardly any evidence for a decelerating phase in the past 15 Gyrs. All the other  Bayes factors in the table are  slightly greater than unity and this shows that if at all there is evidence, it is in favor of an always accelerating universe during this period. However, in sharp contrast to the  model comparison exercise using data of light deflection by sun (discussed in sec. 2), here the evidences are too weak  and hence it is safer to conclude  that the data are unable to discriminate  these two hypotheses.

\ifthenelse{\boolean{ispreprint}}{
\begin{figure*}[p]
\begin{lrbox}{\thisbox}
\epsfig{file=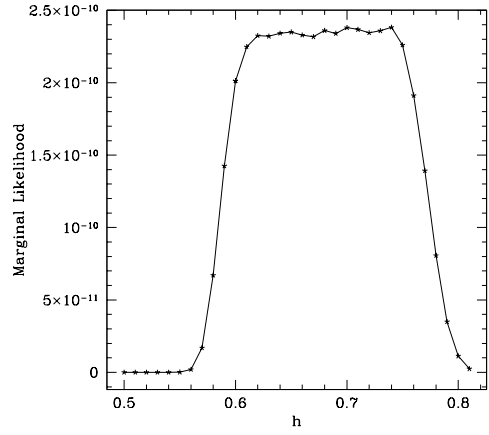, width=3.26in}
\end{lrbox}
\settowidth{\thiswid}{\usebox{\thisbox}}
\begin{center}
\usebox{\thisbox}
\caption{The marginal likelihood for $h$.}
\label{fig:hlik}
\end{center}
\end{figure*}
}{\placefigure{fig:hlik}}

\ifthenelse{\boolean{ispreprint}}{
\begin{figure*}[p]
\begin{lrbox}{\thisbox}
\epsfig{file=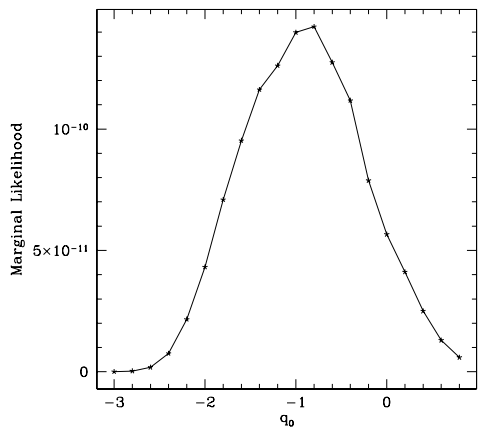, width=3.26in}
\end{lrbox}
\settowidth{\thiswid}{\usebox{\thisbox}}
\begin{center}
\usebox{\thisbox}
\caption{The marginal likelihood for $q_0$.}
\label{fig:qlik}
\end{center}
\end{figure*}
}{\placefigure{fig:qlik}}

\ifthenelse{\boolean{ispreprint}}{
\begin{figure*}[p]
\begin{lrbox}{\thisbox}
\epsfig{file=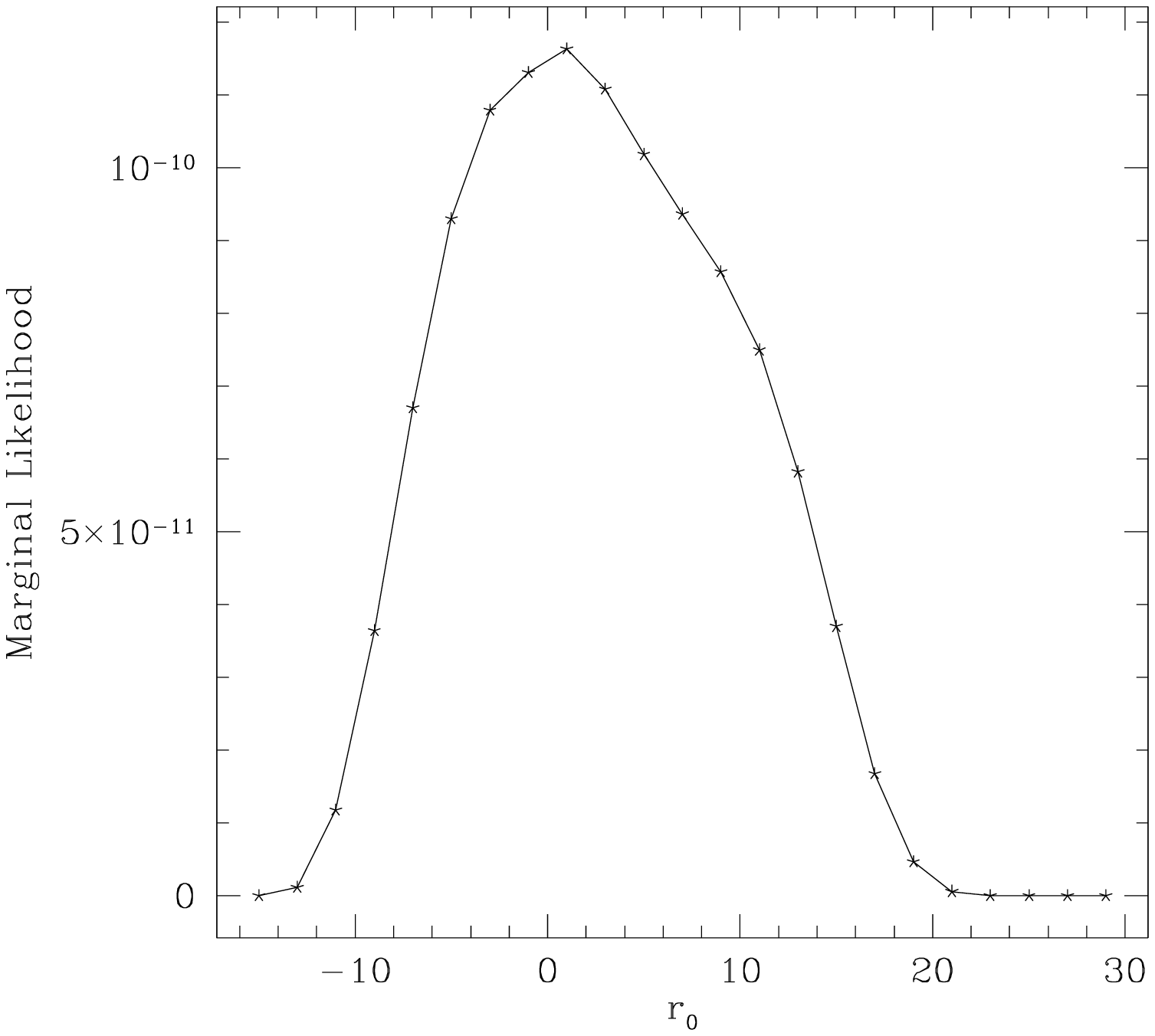, width=3.26in}
\end{lrbox}
\settowidth{\thiswid}{\usebox{\thisbox}}
\begin{center}
\usebox{\thisbox}
\caption{The marginal likelihood for the parameter $r_0$.}
\label{fig:rlik}
\end{center}
\end{figure*}
}{\placefigure{fig:rlik}}

\ifthenelse{\boolean{ispreprint}}{
\begin{figure*}[p]
\begin{lrbox}{\thisbox}
\epsfig{file=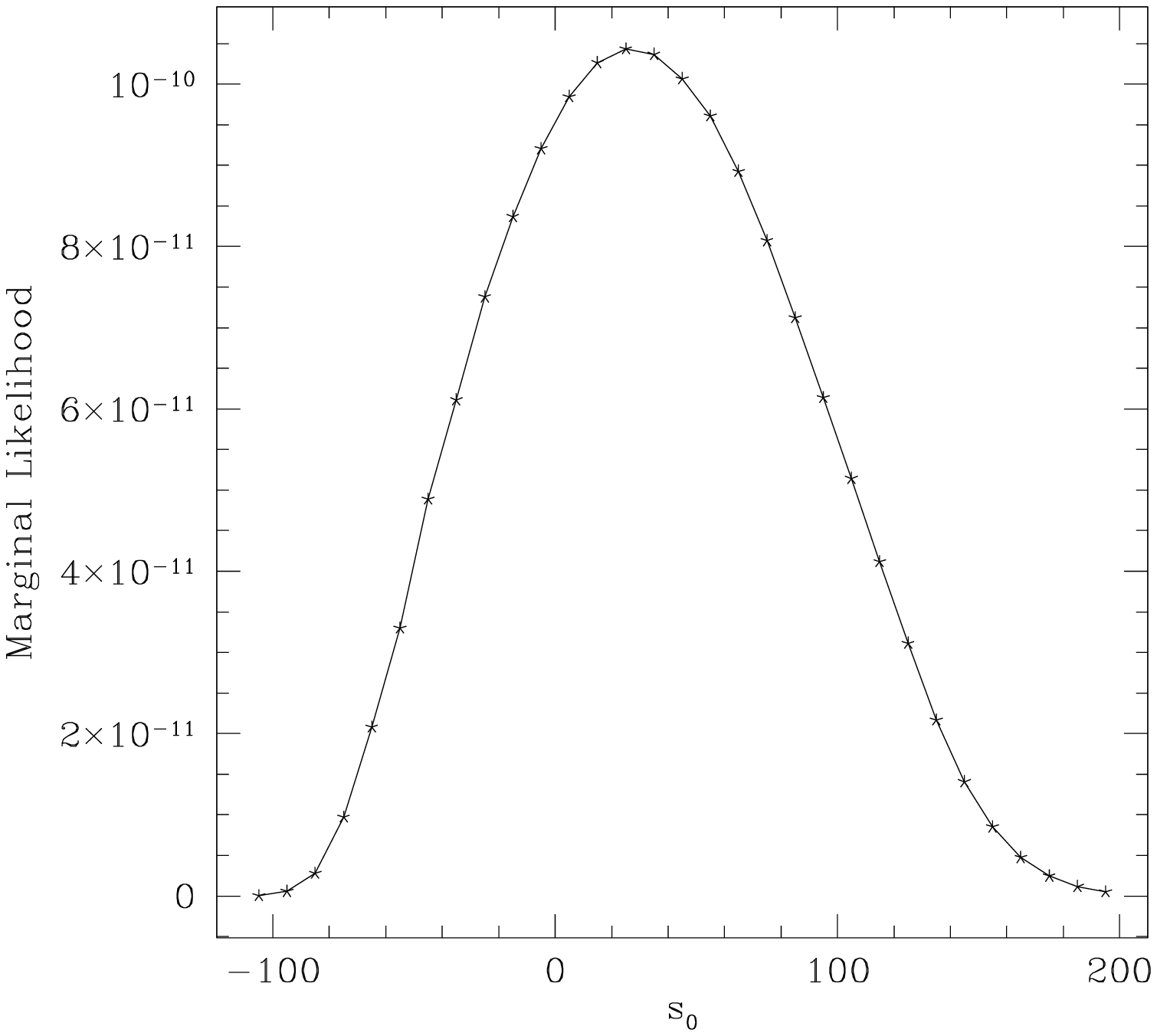, width=3.26in}
\end{lrbox}
\settowidth{\thiswid}{\usebox{\thisbox}}
\begin{center}
\usebox{\thisbox}
\caption{The marginal likelihood for the parameter $s_0$.}
\label{fig:slik}
\end{center}
\end{figure*}
}{\placefigure{fig:slik}}

\ifthenelse{\boolean{ispreprint}}{
\begin{figure*}[p]
\begin{lrbox}{\thisbox}
\epsfig{file=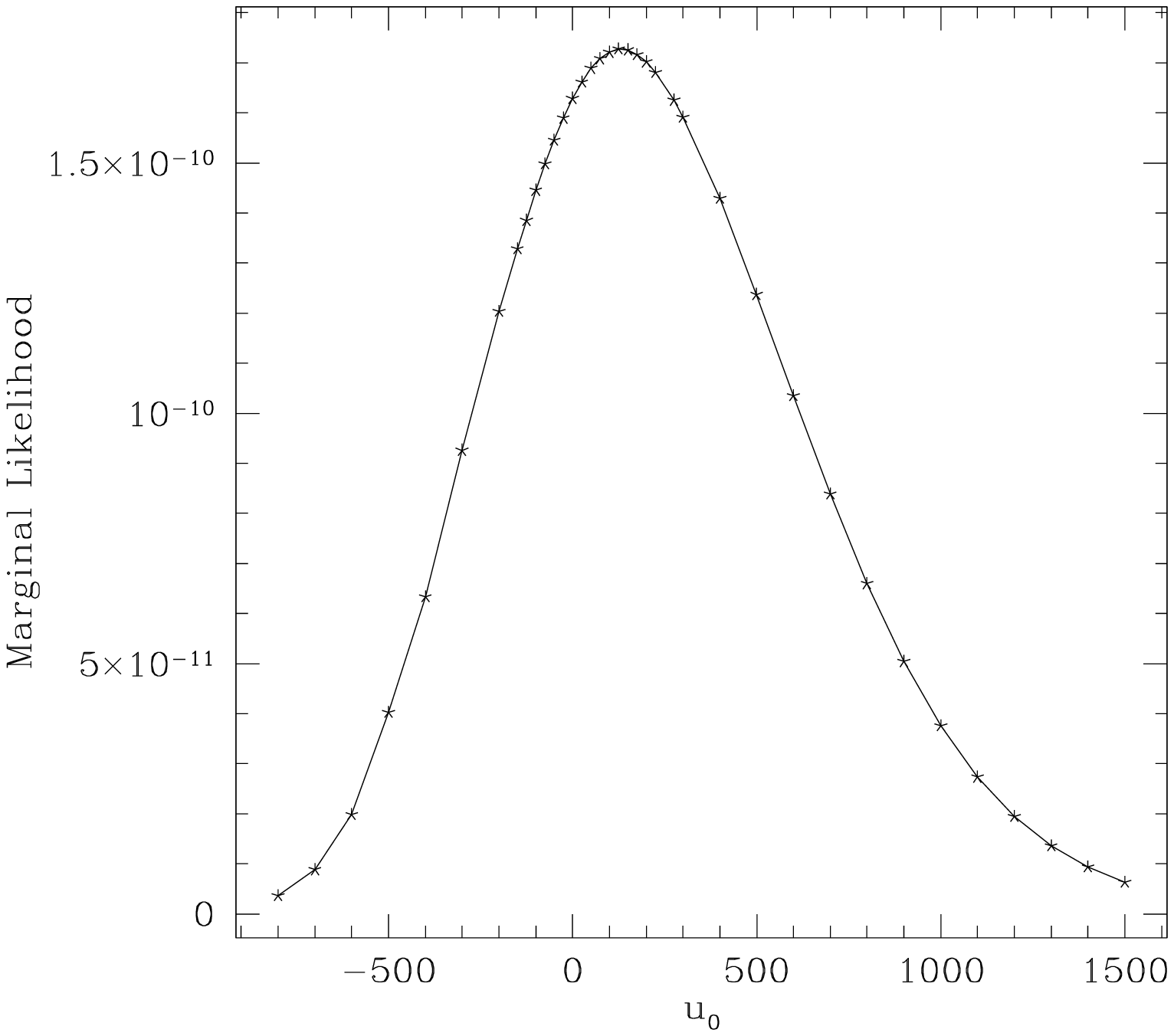, width=3.26in}
\end{lrbox}
\settowidth{\thiswid}{\usebox{\thisbox}}
\begin{center}
\usebox{\thisbox}
\caption{The marginal likelihood for the parameter $u_0$.}
\label{fig:ulik}
\end{center}
\end{figure*}
}{\placefigure{fig:ulik}}

\section{COMPARISON OF SOME FRIEDMANN MODELS}

In this section, we use the method of model comparison to some Friedmann cosmological models. The popular models compared are (1) the Friedmann-Lamaitre-Robertson-Walker  (FLRW) model (Model $M_1$) having a Robertson-Walker (RW) metric with $k=\pm 1,0$, which contains ordinary matter and a ``constant" cosmological constant   and (2) the same FLRW model but with inflation, which implies a $k=0$ RW metric (Model $M_2$). A comparison between these two models itself is an interesting problem,  even though the SNe data do not contain  direct imprints of inflation, except through its prediction of flat space sections or the equivalent condition $\Omega_m+\Omega_{\Lambda}=1$. 

Another model (Model $M_3$) we compare with the above  is a Friedmann model having the simplest evolution with time; i.e., the one whose scale factor $a(t)\propto t$ throughout the history of the universe. This kind of a coasting evolution can be arrived at in a number of ways. The quantity $\rho +3p$, where we denote  the total energy density as $\rho$ and  the total pressure of the universe as $p$, is sometimes called the gravitational charge. In all Friedmann models, if the gravitational charge vanishes, the evolution is coasting. Another approach is to generalize a dimensional argument by Chen and Wu to deduce that the energy densities in the universe, which  are not separately conserved,  vary as $a^{-2}(t)$ \citep{jj}.    In this case too,  a coasting evolution follows. In the above paper, it was noted that when we consider the universe as containing more than one component, eg., matter/radiation and $\Lambda $, the coasting model has a unique feature in which  it predicts a ratio between the density parameters; $\Omega_{m}/\Omega_{\Lambda} =2/(1+3w)$ ($w=0$ or 1/3). Hence it has no  coincidence problem.   It was  shown that this model is devoid of other problems like the horizon,
flatness, monopole,  size, age of the universe and the generation of density perturbations on scales well above the present Hubble radius in the classical epoch. An added advantage is that here one can consider the generation of the observed density perturbations as a late-time classical behavior too. The solution of the coincidence, age and density
perturbation problems deserve special mention since these problems are
not solvable in an inflationary scenario.  It was also shown
that the evolution of temperature in the model is nearly the
same as that in the standard big bang model  and this will enable nucleosynthesis to proceed in
an identical manner, provided the total density parameter $\Omega \approx 2$. In \cite{mvjvn}, the afore-said condition was stated to be a problem with model $M_3$.

The above coasting evolution  can also be obtained by a quite different route \citep{jj1,jj2}. Here we consider the analytic continuation of a real analytic manifold (the spacetime) into the complex,
producing a complex spacetime.  There are a number of instances of the use of complex numbers or complex analytic functions in general theory of relativity \citep{fla,newmann}. 
This model is arrived at by extending the idea of a possible `signature change' in
the  universe \citep{ellis,senovil}, a widely discussed speculation which involves
some basic issues in the general  relativity. This extension leaves us in an
unphysical universe, but it was noticed that a proper interpretation
of the theory will enable us to obtain a cosmological
model, with the essential features as summarized above. The evolution of $a(t)$ is obtained as $a^{2}(t)=a_{min}^2+c^2t^2$ and a quantum cosmological calculation shows that $a_{min} \sim l_p$, the Planck length. Thus after the Planck epoch, $a\propto t$ and the evolution is the same as in the above case. In addition to the above advantages, this approach solves the singularity problem too.

However, the analysis in this paper is intended to provide and improve an example of Bayesian model comparison \citep{mvjvn} and the coasting model is  suitable for this purpose due to its simplicity. This model is  not a realistic one since it  requires some more fundamental derivation such as that in the scalar field dark energy models. As in the previous work, we assume that the model is given and then proceed with the evaluation. 

One observes that when the Bayes factor is near unity, the prior odds $p(M_i|I)/p(M_j|I)$ in equation (\ref{eq:odds}) becomes very important. While comparing models $M_1$, $M_2$, and $M_3$, it was observed in the above paper that whereas models $M_1$ and $M_2$ are plagued by the large number of cosmological  problems,   $M_3$ suffers from the heuristic nature of its derivation and the problem with nucleosynthesis, as mentioned earlier.  But one can argue that the inclusion of a small $\Lambda $ in the present universe in models $M_1$ and $M_2$ itself is done heuristically.  However, a balance was set between the prior probabilities and  these models were compared with prior odds equal to unity. The results thus obtained in the previous work,  while using the $m-z$ data \citep{per99} for comparison were $B_{13} \approx 3.1$ and $B_{23} \approx 5$ and the conclusion made was that  there is some but  not strong evidence against model $M_3$.

The data used in the present analysis are the same as those used in the calculations in the previous section.
The parameters in the models are $H_0$, $M$, $\Omega_m$ and $\Omega_{\Lambda}$. But in $M_2$ and $M_3$, their numbers are effectively reduced by one each, due to the conditions $\Omega_m+\Omega_{\Lambda}=1$ and $\Omega_{m}/\Omega_{\Lambda} =2/(1+3w)$ ($w=0$ or 1/3), respectively, in these models. While using data $D_1$, one can consider  $H_0$ and $M$ as a single parameter by  suitably combining them, but while using data $D_2$, we do not need to include $M$. However, in the present analysis using $D_1$, we consider $H_0$ and $M$ as independent parameters, as we did in section 3 and they are assumed to have flat priors in the ranges  $0.6<h<0.8$ and $-19.6<M<-19.1$, respectively.  This is a modification of the procedure in \cite{mvjvn}, but this alone does not affect the results in any  way. 
  Since the data  consist of several common SNe and many of them are refined values of those used in the above paper, we decide to make their analysis independently of the  one in that work. However, the information $I$ we have regarding the  ranges of $\Omega_{m}$ and $\Omega_{\Lambda}$ are modified on the basis of the previous results. The new ranges  chosen are $0<\Omega_{m}<1$ and $0<\Omega_{\Lambda}<1$. One can see that small variations in these ranges do not affect our conclusions drastically.

The luminosity distance $D$ in these cases can generally be written as $D=(c/H_0)d(z; \Omega_{m},\Omega_{\Lambda})$. Thus the apparent magnitude is $m=5\log (c/H_0) + 5\log d(z; \Omega_{m},\Omega_{\Lambda})+25+M$. The expression for $d$ to be used for $M_1$ and $M_2$ is

$$ 
d=5\log \{ (1+z)\; |\Omega_k |^{-1/2}
S_k[|\Omega_k |^{1/2} I(z)]\} , 
$$
where    $\Omega_k =1-\Omega_m -\Omega_{\Lambda }$ and $S_k(x) =\sin x$ for $\Omega_m + \Omega_{\Lambda }>1$, $S_k(x) =\sinh x$ for $\Omega_m + \Omega_{\Lambda }<1$  and
$S_k(x) = x$ for $\Omega_m + \Omega_{\Lambda }=1$. Also

$$ 
I(z) = \int _{0}^{z} [(1+z^{\prime })^{2}(1+\Omega_{m}z^{\prime})-z^{\prime}(2+z^{\prime} )(\Omega _{\Lambda})]^{-1/2}\;  dz^{\prime}. 
$$
For Model 3, the function $d$  can be written as

$$ 
d=5 \log \{ m(1+z)S_k[\frac {1}{m}\ln (1+z)] \} , 
$$
where $m=\sqrt {2k/(3\Omega _{m} -2)}$ for the nonrelativistic era and $S_k(x) =\sin x$ for $\Omega_m >2/3$, $S_k(x) =\sinh x$ for $\Omega_m <2/3$  and $S_k(x) = x$ for $\Omega_m =2/3$.

 The Bayes factors are evaluated as in the previous section and the results obtained in the present analysis are tabulated in Table 2. These  show that the FLRW and inflationary models are at a  more advantageous position than in  \cite{mvjvn}, as per the interpretation of the Bayes factor.

\section{DISCUSSION}

Our attempt in this paper is to apply the Bayesian model comparison method to some vexing problems in cosmology in the light of latest $m-z$ data of SNe Ia. These data are the results of potential landmark observations in cosmology, after the discovery of the Hubble's law.  Since they  point to an accelerating universe today, it is natural to ask for how long the universe remained in this phase.  The Bayesian analysis shows that  except in one case, there is no evidence from SNe data to conclude that a changeover from deceleration to acceleration occurred anywhere in the past $5\times 10^{17}$ s.

But a rider we add to the above conclusion is even more important. The odds ratios between `always accelerating' and `decelerating' models in the given period, with impartial prior odds, are obtained to be  close to unity,  and that too in favor of the former model in most cases. However, as per the interpretation of this ratio, it is highly objectionable to state that one or the other of the models analyzed is ruled out, when it is on the basis of obtaining this ratio close to unity. To highlight this point,  the model comparison exercise was performed to one of the classic tests of general relativity, viz., the deflection of light by sun, as it just grazes the sun's surface. The huge value of the odds ratio  (more strictly, the Bayes factor, since   the prior odds was taken to be unity) obtained tells us how the Newtonian explanation of this phenomenon is ruled out by this observation. One can see that epoch-making discoveries are always accompanied by such large values for the Bayes factor.  We wish anybody who claims that some particular model is `ruled out' by some data to take note of this fact.  Bayesian theory also teaches the important lesson that the price for overlooking this fact would be unavoidable and frequent U-turns.  The present analysis rules out neither the accelerating nor the decelerating models; instead, we safely conclude that the data cannot discriminate these models.

The present analysis makes use of a significant improvement in the cosmographic approach adopted in \cite{mvj}. In order to overcome the problem of truncations in the series for $1/a(t)$ used in the equation $1+z=a_0/a(t)$ in the previous work,  the solution of this equation is performed here in a purely numerical way. Though this consumes more computation time, it is advantageous that the accuracy is not compromised. As a consequence of the above modification, we had to perform a numerical integration in the expression for $r_1$. Though these amount to application of brute force in the analysis, the procedure has become much transparent. When compared to our own and other model-independent analyses of SNe data carried out previously    (mentioned in the introduction), the present one is simpler and  uses the least amount of clumsy formulas and this makes the treatment more intelligible. The procedure also makes use of the marginal likelihood method in the analysis. Since we are basically interested in the coefficients in the series expansion and the odds ratio, it is better to keep the treatment simple, to ensure accuracy at every stage. Considerable enhancement in such accuracy can be noticed from the plots of new marginal likelihoods for $h$, $q_0$, $r_0$ $s_0$,and $u_0$ (shown in Figs. 1-5, respectively, which are drawn using the  data $D_1$), when compared with the corresponding curves in \cite{mvj}.

It may be recalled that the likelihood for the model ${\cal L}(M_i)$  computed using equation (\ref{eq:likelimod}) is in fact  the probability for the data, given the model and $I$ are true and is not exactly the `probability for the model'. Similarly, in the Bayesian scheme, the marginal likelihood for a parameter is not to be considered as the probability distribution for the parameter. However, we have computed the mean and $\sigma$ values from the above marginal likelihoods. Their new values are $h=0.68\pm 0.06$, $q_0=-0.90\pm 0.65$, $r_0=2.7\pm 6.7$, $s_0=36.5\pm 52.9 $, and $u_0=142.7\pm 320$.       The large $\sigma$-values show that indeed there are a  variety of choices for the values of the parameters, which may be considered as `good fit' in the conventional way.

As stated earlier, the $\chi^2$-values and hence the Bayes factor crucially depend on the error bars. The fact that the data cannot clearly discriminate different models in cosmology implies that the error bars in the data are quite large. Hopefully, future observations will have sufficiently small errors so that the Bayes factor between different models may become large. But some important points one should check in such cases is whether the errors are truly Gaussianly distributed and also whether we know accurately the standard deviation of these errors. Any deviation from these conditions will  affect the validity of the model comparison. In cases where we are not sure about these two assumptions, one can  resort to some other version of probability theory; for instance, the median statistics advocated by \cite{gott} for analyzing cosmological datasets. These examples demonstrate how important are detailed considerations of the underlying assumptions in making judgments on what the data tells in cosmology.

It should be admitted that the Taylor expansion of the scale factor  has more parameters to be constrained than realistic models and thereby it weakens the power of the data.  But the cosmographic approach, which is basically a kinematic one, is not an alternative to  realistic cosmological models. Instead, it helps to consolidate the evidences for such  models. For example, evaluating the expansion parameters like $h$, $q_0$, $r_0$, $s_0$, etc. will help to constrain these models, as in the traditional use of the values of $h$ and $q_0$.  It is also of use in answering questions like how long into the past the accelerating phase prevailed, as we have attempted to do in this paper.  When we have more terms in the Taylor expansion, our extrapolation into the past becomes more accurate. Truncating at the fifth-order is due to practical considerations. Adding one more term would increase the computation time at least by an order of magnitude. We have seen that for extrapolating to the past corresponding to $z \approx 1.75$,  up to fifth-order term in the expansion is required.

In this paper, we have also performed the  comparison between some Friedmann models such as FLRW, inflationary and coasting models. This is viewed as an extension of a previous work using new and refined data. This analysis uses the previous one in modifying the flat priors. We use flat priors for the parameters $\Omega_m$ and $\Omega_{\Lambda}$ in the intervals $0<\Omega_m<1$ and $0<\Omega_{\Lambda}<1$. The Bayes factors show that when data $D_1$ is used, the  FLRW and inflationary models are at par with each other, and there is positive evidence against the coasting models. But in the case of using $D_2$, when compared to the the first model, there is some evidence against the inflationary one and a strong evidence against the coasting one. Also when compared to inflationary models, there is a positive evidence against the coasting model.  But we again recall that these evidences are not very strong  to rule out any of them.
 It is also to be noted that there is no contradiction between the two analyses in sections 3 and 4. The results in the latter section simply states that when compared to   models $M_1$ and $M_2$, which are largely accelerating, the non-accelerating model $M_3$ is disfavored. Thus the evaluation of Bayes factor helps  to quantify our knowledge, even when we are not aware of the full story.

These lessons are important in keeping cosmology a science. So far in the history of physics, though there were scientific revolutions, we always find that the new theories contain the old ones as limiting cases. This is because experimental evidences do not contradict the older theories  in the realms in which they are applicable. For example, in a laboratory experiment where general relativistic effects are negligible, the Bayes factor between Newtonian and Einsteinian theories will be close to unity. Our results show that cosmological observations have not yet passed this stage and it is our opinion that  the claims in cosmology, including the old and new ones we have compared in this paper, are not convincing enough since they are not  supported by satisfactorily large  Bayes factors. In physics we also see that in the realms where new theories are indispensable, they make unambiguous predictions, which are then verified experimentally.  It can be seen that in such cases, new theories are often supported by huge Bayes factors, as in the example of light deflection discussed above in this paper.   To achieve such goals in cosmology, we should tread with caution, begin with the cosmographic approach since it is the most fundamental one, and   Bayes theory will help us to look for hard evidences. We have made such an attempt in this paper and the main conclusion, which is consistent with other approaches, is that the available data cannot clearly discriminate the cosmological models analyzed.

\acknowledgments

The author wishes to thank the UGC for a Research Grant and  IUCAA, Pune, where part of this work was done, for hospitality.

\begin{table}[H]
\scriptsize\renewcommand{\arraystretch}{1.0}
\begin{lrbox}{\thisbox}
\begin{tabular}{ccc}
\tableline
\tableline
$T_{p}$ s &  $B_{12}$ using Data $D_1$ &  $B_{12}$ using Data $D_2$ \\
\tableline
$-1\times 10^{17}$ &3.3 & 1.1 \\
$-2\times 10^{17}$ &1.5 & 0.6 \\
$-3\times 10^{17}$ &2  & 1.1 \\
$-4\times 10^{17}$ &2.4 & 1.6 \\
$-5\times 10^{17}$ &2.8  & 2.1\\
\tableline
\end{tabular}
\end{lrbox}
\settowidth{\thiswid}{\usebox{\thisbox}}
\begin{center}
\begin{minipage}{\thiswid}
\caption{Bayes Factors between "Always Accelerating" and "Decelerating" Cosmological Models during the past. }
\label{tab:alacdec}
\usebox{\thisbox}

\end{minipage}
\end{center}
\end{table}

\begin{table}[H]
\scriptsize\renewcommand{\arraystretch}{1.0}
\begin{lrbox}{\thisbox}
\begin{tabular}{ccc}
\tableline
\tableline
Bayes Factor &  Using Data $D_1$ &  Using Data $D_2$ \\
\tableline
$B_{12}$ &0.9 &1.8  \\
$B_{13}$ &12 & 27.5 \\
$B_{23}$ &13.4 & 15.3 \\
\tableline
\end{tabular}
\end{lrbox}
\settowidth{\thiswid}{\usebox{\thisbox}}
\begin{center}
\begin{minipage}{\thiswid}
\caption{Bayes Factors between Friedmann Models $M_1^a$, $M_2^b$, and $M_3^c$. }
\label{tab:friedmann}
\usebox{\thisbox}
a: $M_1$- FLRW model with matter and a cosmological constant.

b: $M_2$- The same as $M_1$, but with inflation ($\Omega_m+\Omega_{\Lambda}=1$).

c: $M_3$- Coasting model.

\end{minipage}
\end{center}
\end{table}


\end{document}